\newcommand{\href}[2]{{\bf #2} (\texttt{#1})}

 \documentstyle[11pt,aaspp]{article}

\begin{document}

\title{\Large\bf Cosmological Hydrodynamics with Multi-Species Chemistry
and Nonequilibrium Ionization and Cooling}

\author{\large Peter Anninos, Yu Zhang, Tom Abel and Michael L. Norman}
\vskip12pt
\affil{\it Laboratory for Computational Astrophysics \\
       National Center for Supercomputing Applications \\
       University of Illinois at Urbana-Champaign \\
       405 N. Mathews Ave., Urbana, IL 61801}


\begin{abstract}
We have developed a method of solving for multi-species chemical reaction flows
in non--equilibrium and self--consistently with the hydrodynamic equations
in an expanding FLRW universe. The method is based
on a backward differencing scheme for the required stability when solving
stiff sets of equations and is designed
to be efficient for three-dimensional calculations without
sacrificing accuracy.
In all, 28 kinetic
reactions are solved including both collisional and radiative processes for the
following nine separate species:
$H$, $H^+$, $He$, $He^+$, $He^{++}$, $H^-$, $H_2^+$, $H_2$, and $e^-$.
The method identifies those reactions (involving
$H^-$ and $H_2^+$) ocurring on the shortest time scales, decoupling them from
the rest of the network and
imposing equilibrium concentrations to good accuracy
over typical cosmological dynamical times.
Several tests of our code
are presented, including radiative shock waves, cosmological sheets,
conservation constraints, and fully three-dimensional simulations of
CDM cosmological evolutions in which we compare our method to results
obtained when the packaged routine LSODAR is substituted for our algorithms.
\end{abstract}


\section{Introduction}
\label{sec:introduction}

Hydrodynamical and microphysical processes of baryonic matter play an important
role in structure formation at the smaller sub--cluster scales. Indeed,
microphysics can play a more important role than gravity, especially when
the cooling time of the gas is much shorter than the dynamical or Hubble times.
Protogalactic clouds, which may form, for example, at high redshifts in
CDM models can collapse through cooling instabilities to form an early
generation of stars. The feedback from primordial stars can change
the physical state of the pre--galactic medium and thus have considerable
influence over the subsequent formation of stars and galaxies and the
general state of the intergalactic medium
(Couchman \& Rees 1986; Tegmark et al. 1996).
Microphysical processes are also very important at the center--most regions of
cosmological sheets or pancakes. Originally studied by Zel'dovich (1970)
in the context of neutrino-dominated cosmologies, sheets are ubiquitous
features in nonlinear structure formation simulations of CDM-like models
with gas, and manifest on a spectrum of length scales and formation epochs.
Cooling processes occur on very short time scales at the center--most
densest parts of the pancake structures where stars and galaxies can form
from the fragmentation of the gas.
(Bond et al. 1984; Shapiro \& Struck--Marcell 1985;
Yuan, Centrella \& Norman 1991; Anninos \& Norman 1996).

It is well known that when nonequilibrium atomic reactions are properly
taken into account, the cooling time can be shorter than the hydrogen
recombination time and that gas which cools to $1~eV$ will likely
cool faster than it can recombine.
The effect of this nonequilibrium
cooling is to leave behind a greater residual of free electrons and ions,
as compared to the equilibrium case. The free electrons can be captured
by neutral hydrogen to form $H^-$ that subsequently produce hydrogen molecules.
If large concentrations of molecules can form, the cooling is
dominated by the vibrational/rotational
modes of molecular hydrogen which acts to efficiently cool
the gas to about $10^{-2}~eV$, thereby reducing the Jeans mass of the gas.
Hydrogen molecules can therefore play a crucial role in the formation
of stars as they provide the means for cloud fragments to collapse and
dissipate their energy.

However, the typically high computational requirements and technical
difficulties
needed to solve the chemical rate equations relevant for $H_2$
production in hydrodynamic flows have forced previous
authors to impose simplifying assumptions such as the steady
state shock condition which reduces the problem to zero
dimension. In this case, only the time development of the hydrodynamic
and thermodynamic variables are solved
(Izotov \& Kolesnick 1984; MacLow \& Shull 1986; Shapiro \& Kang 1987;
Kang \& Shapiro 1992).
All of these studies have consistently found that the mass fraction
of $H_2$ can reach $10^{-3}$ behind sufficiently strong shocks,
which is adequate to cool the gas
to temperatures of order $10^{-2}~eV$ well within a Hubble time.
More recently Haiman, Thoul \& Loeb (1995) have investigated the formation 
of low mass objects in one--dimensional numerical calculations.
They confirm the importance of $H_2$ in the collapse of
spherically symmetric isolated objects at high redshifts.

Although much insight has been gained about the chemical aspects of
molecular hydrogen formation and cooling, it remains to incorporate chemical
reaction flows in realistic cosmological models.
This paper discusses a method that we have developed for solving the
kinetic rate equations with multi--species chemistry in
nonequilibrium and self--consistently with the hydrodynamic
and $N$--body equations in an expanding FLRW universe. The method is based
on a backward differencing formula (BDF) for the required stability
when solving stiff sets of equations, and
is designed for both accuracy but especially speed
so that it may be used in three dimensional codes with a minimal
strain on computational resources.
In all, we solve for 28 kinetic reactions
including collisional and radiative processes for nine different
species: $H,~H^+,~He,~He^+,~He^{++},~H^-,~H_2^+,~H_2$, and $e^-$,
which we track individually with their unique mass transport equations.
We have also implemented a comprehensive model
for the radiative cooling of the gas that includes
atomic line excitation, recombination, collisional ionization,
free-free transitions, molecular line excitations,
and Compton scattering of the cosmic background radiation by electrons.

The set of hydrodynamic, $N$--body, kinetic and cosmological
equations that we solve are summarized in \S \ref{sec:equations}.
Section \ref{sec:numerical} discusses the BDF method and its integration
into the hydrodynamic solver.
Several tests of our code are presented in \S \ref{sec:codetests},
including 1D radiative shock waves, 2D cosmological sheets,
and fully 3D simulations of
CDM cosmological evolutions in which we compare the BDF method to results
obtained when the packaged routine LSODAR 
(Hindmarsh 1983; Petzold 1983) is substituted in its place.
We provide concluding remarks in \S \ref{sec:summary}.

Finally we note that a companion paper has been written which discusses the
chemical model and cooling functions in more detail (Abel et al. 1996).
In that paper, we motivate the model and argue its comprehensive nature in the
choice of reactions, insofar as the formation of hydrogen molecules
in cosmological environments is concerned. We also provide more up--to--date
and accurate fits to the different rate coefficients.
For completeness, we tabulate the reaction list and cooling
processes in Appendices A and B of this paper, but leave the rate coefficients
to Abel et al. (1996).

\section{The Equations}
\label{sec:equations}

The hydrodynamical equations for mass, momentum and energy conservation
in an expanding FRW universe with comoving coordinates are
\begin{equation}
  \frac{\partial {\rho_b}}{\partial t} + \vec{\nabla} \cdot
({\rho_b}{\vec{v}_b})
    + 3{\frac{\dot a}{a}}{\rho_b} = 0 ,
\label{hydromass}
\end{equation}
\begin{equation}
  \frac{\partial ({\rho_b}{{v}_{b,i}})}{\partial t}
    + \vec{\nabla} \cdot [({\rho_b}{{v}_{b,i}}){\vec{v}_b}]
    + 5{\frac{\dot{a}}{a}}{\rho_b}{{v}_{b,i}}
    = - {\frac{1}{a^2}}{\frac{\partial p}{\partial x_i}}
      - {\frac{\rho_b}{a^2}}{\frac{\partial \phi}{\partial x_i}} ,
\label{hydromom}
\end{equation}
\begin{equation}
  \frac{\partial e}{\partial t} + \vec{\nabla} \cdot (e{\vec{v}_b})
    + p\vec{\nabla} \cdot {\vec{v}_b} + 3\frac{\dot a}{a}(e+p)
    = \Gamma - {\dot E} ,
\label{hydroenergy}
\end{equation}
where $\rho_b$, $p$ and $e$ are the baryonic density, pressure and specific
internal energy defined in the proper reference frame, $\vec{v}_b$ is the
comoving peculiar baryonic velocity, $\phi$ is the
comoving gravitational potential that
includes baryonic plus dark matter contributions, $a=1/(1+z)$ is the
cosmological scale factor, and $\dot E$ and $\Gamma$ are the
microphysical cooling and heating rates.

The equations for collisionless dark matter in comoving coordinates are
\begin{eqnarray}
   \frac{d{\vec{x}_d}}{dt} &=& {\vec{v}_d} ,  \\
   \frac{d{\vec{v}_d}}{dt} &=& - 2{\frac{\dot{a}}{a}}{\vec{v}_d}
                                  - {\frac{1}{a^2}}{\vec{\nabla}}{\phi} .
\end{eqnarray}
The baryonic and dark matter components are coupled through Poisson's equation
for the gravitational potential
\begin{equation}
   {{\nabla}^2}{\phi} = 4{\pi}G{a^2}({\rho}-\bar{\rho}) ,
\end{equation}
where $\rho = \rho_b + \rho_d$ is the total density and
$\bar{\rho}=3H_0^2\Omega_0/(8\pi G a^3)$
is the proper background density of the universe.

The cosmological scale factor $a(t)$ is given by Einstein's equation
\begin{equation}
   \frac{d a}{dt} = H_0 \left[ \Omega_M (\frac{1}{a} -1)
                               +\Omega_\Lambda (a^2 -1) + 1 \right]^{1/2}
\end{equation}
where $\Omega_M=\Omega_b + \Omega_{d}$ is the density parameter including
both baryonic and dark matter contributions,
$\Omega_\Lambda= \Lambda /(3H_0^2)$ is the
density parameter attributed to the cosmological constant $\Lambda$,
and $H_0$ is the present Hubble constant.

In addition to the usual hydrodynamic equations (\ref{hydromass}) --
(\ref{hydroenergy}), we must also solve equivalent mass conservation
equations for the densities $\rho_i$ of each of the nine separate
atomic and molecular species that we track
\begin{equation}
  \frac{\partial {\rho_i}}{\partial t} + \vec{\nabla} \cdot
({\rho_i}{\vec{v}_b})
    + 3{\frac{\dot a}{a}}{\rho_i} = \pm {\sum_{j}}{\sum_{l}}
                                        {{k_{jl}(T)}{\rho_j}{\rho_l}}
                                    \pm \sum_{j} I_j \rho_j ,
\label{hydrospecies}
\end{equation}
where the signs of each term on the right--hand--side depend on
whether the process creates or destroys the species $\rho_i$.
The ${k_{jl}(T)}$ are rate coefficients for the two body reactions
and are functions of the gas temperature $T$.
Explicit analytic fits for these coefficients over a broad range of
temperatures,
and a general discussion of the relevant chemical reactions,
can be found in Abel et al. (1996).
In all, we include 28 rate coefficients, one for each of the
chemical reactions shown in Appendix A.
The $I_j$ in equation (\ref{hydrospecies}) are integrals due
to photoionizations and photodissociations
\begin{equation}
  I_j = \int_{\nu_{0,j}}^{\infty} 4\pi\sigma_j(\nu)
             \frac{{\cal I}(\nu)}{h\nu} d\nu ,
\end{equation}
where ${\cal I}(\nu) = {\cal F}(\nu)/4\pi$ is
the intensity of the radiation field, ${\cal F}(\nu)$ is the flux,
$\sigma_j (\nu)$ are the cross-sections for the
photoionization and photodissociation processes,
and $\nu_{0,j}$ are the frequency thresholds for the respective processes.
We note that the nine equations represented by
(\ref{hydrospecies}) are not all independent.
The baryonic matter is composed of hydrogen and helium with a fixed
primordial hydrogen mass fraction of $f_H$. Hence we have the following
three conservation equations
\begin{eqnarray}
  \mbox{hydrogen nuclei:}&& \qquad
    \rho_H + \rho_{H^+} + \rho_{H^-} + \rho_{H_2^+} + \rho_{H_2} = \rho_b~f_H ,
    \label{conserveH} \\
  \mbox{helium nuclei:}&& \qquad
    \rho_{He} + \rho_{He^+} + \rho_{He^{++}} = \rho_b~(1-f_H) ,
    \label{conserveHe} \\
  \mbox{charge conservation:}&& \qquad
    \rho_{H^+} - \rho_{H^-} + \frac{1}{2}\rho_{H_2^+} +
    \frac{1}{4}\rho_{He^+} + \frac{1}{2}\rho_{He^{++}} = m_H~n_e ,
    \label{conserveNe}
\end{eqnarray}
where $n_e$ is the number density of free electrons and $m_H$ the
proton mass.

To complete the set of equations (\ref{hydromass}) -- (\ref{hydroenergy}),
we must also specify the equation of state appropriate for an ideal gas
\begin{equation}
  e = \frac{p}{\gamma-1} = \frac{k_B T}{\gamma-1} \sum_{i=1}^{9} n_i ,
\label{eqos}
\end{equation}
where $\gamma=5/3$ is the ratio of specific heats for the baryonic
matter, $k_B$ is boltzmann's constant, $T$ is the gas temperature,
and $n_i$ are the number densities for each of the different species.
We also need to provide the necessary cooling $\dot E$ and heating
$\Gamma$ functions
to the right-hand-side of equation (\ref{hydroenergy})
\begin{eqnarray}
  \dot E &=& \dot E_{Comp} +  \sum_{j} \sum_{l}
                       \dot e_{jl}(T) \rho_j \rho_l , \\
  \Gamma &=& \sum_{j} J_j(\nu) \rho_j ,
\end{eqnarray}
where $\dot E_{Comp}$ is the Compton cooling (or heating) due to
interactions of free electrons with the cosmic microwave
background radiation, and
$\dot e_{jl}(T)$ are the cooling rates from two-body
interactions between species $j$ and $l$.
The ${J_j(\nu)}$ are integrals due to photoionizing and
photodissociating heating
\begin{equation}
  J_j(\nu) = \int_{\nu_{0,j}}^{\infty} 4\pi \sigma_j(\nu) {\cal I}(\nu)
             \frac{(h \nu - h \nu_{0,j})}{h\nu} d\nu .
\end{equation}
We include a total of fourteen processes in the
cooling function and three processes for heating.
The physical mechanisms and mathematical expressions for each process
are given in Appendix B.

\section{NUMERICAL METHODS}
\label{sec:numerical}

It is well known that the differential equations describing
non-equilibrium atomic and molecular rate reactions can exhibit
variations on extreme time scales. Characteristic creation and destruction
scales can differ by many orders of magnitude among the different species
and reactions. As a result, explicit schemes for integration can be unstable
unless unreasonably small time steps (smaller than the shortest
dynamical times in the reaction flow) are taken, which makes any
multi-dimensional
computation prohibitively expensive.  For this reason implicit
methods are preferred for stiff sets of equations. These methods
generally involve a Newton's iterative procedure to achieve convergence, and
for large dimensional Jacobian matrices these implicit methods can also
be very time consuming. A number of packaged routines exist which are based on
identifying the disparity in time scales among the species and switching
between stiff and nonstiff solvers. An example of such a
package is the Livermore solver for ordinary differential equations
with automatic method switching for stiff and nonstiff problems LSODAR
(Hindmarsh 1983; Petzold 1983). However, an implementation of this
solver in multi-dimensions is extremely costly in computer time and
an alternative numerical scheme is desirable for fully three-dimensional
calculations where computational speed is crucial.

\subsection{General Framework}
\label{subsec:general}

We use an
operator and directional splitting of the hydrodynamic
equations (\ref{hydromass}) -- (\ref{hydroenergy})
and (\ref{hydrospecies}) to update the fourteen
state variables $\rho_b$, $\vec v_b$, $e$ and $\rho_i$.
Six basic steps are utilized.

First, the source step accelerates the fluid velocity
due to pressure gradients and gravity, and modifies the velocity
and energy equations to account for artificial viscosity
\begin{eqnarray}
\rho_b\frac{\partial\vec{v}}{\partial t} &=&
     -\frac{1}{a^2}\vec{\nabla} p -\frac{\rho_b}{a^2}\vec{\nabla}\phi ,
\label{split1}\\
\rho_b\frac{\partial\vec{v}}{\partial t} &=&
     -\vec{\nabla}\cdot Q ,  \\
\frac{\partial e}{\partial t} &=& -Q:\vec{\nabla}\vec{v} ,
\end{eqnarray}
where $Q$ is a second rank tensor representing the artificial viscous
stresses (Stone \& Norman 1992). A staggered mesh scheme is utilized
whereby the scalar variables $\rho_b$, $\rho_i$, $e$, $\phi$ and the artificial
viscosity are zone centered,
while the velocities are located at the zone interfaces. The pressure,
potential, and viscosity gradients are thus naturally aligned with the momentum
terms.

In the second cooling/heating step, the energy changes are computed from
``pdv'' work and radiative cooling and heating from microphysical processes
\begin{equation}
\frac{\partial e}{\partial t} = -p\vec{\nabla}\cdot\vec{v}
                                -\dot E_{cool} + \Gamma .
\label{split2}
\end{equation}
We discuss solving this equation further in the following subsection.

The third expansion step updates all state variables from the terms
arising from the expansion of the universe
\begin{eqnarray}
\frac{\partial\rho_b}{\partial t} &=& -3\frac{\dot a}{a}\rho_b , \\
\frac{\partial\rho_i}{\partial t} &=& -3\frac{\dot a}{a}\rho_i , \\
\frac{\partial\rho_b \vec{v}}{\partial t} &=&
                                      -5\frac{\dot a}{a}\rho_b\vec{v} , \\
\frac{\partial e}{\partial t} &=& -3\frac{\dot a}{a}(e+p) .
\end{eqnarray}
The homogeneous nature of the expansion allows a simple solution
$\rho_b^{t+\Delta t} = (a^t/a^{t+\Delta t})^3 \rho_b^t$,
although a more generalized procedure is required for the energy equation
in which an effective adiabatic index must be defined to eliminate
the pressure term in the case of ionized gas (Anninos \& Norman 1994).

The fourth or transport step solves the advection terms
\begin{eqnarray}
\frac{d}{dt}\int \rho_b dV &=& -\int\rho_b\vec{v}\cdot d\vec{S} , \\
\frac{d}{dt}\int \rho_i dV &=& -\int\rho_i\vec{v}\cdot d\vec{S} , \\
\frac{d}{dt}\int \rho_b v_i dV &=& -\int\rho_b v_i \vec{v}\cdot d\vec{S} , \\
\frac{d}{dt}\int e dV &=& -\int e\vec{v}\cdot d\vec{S} \label{split4}.
\end{eqnarray}
Several different monotonic schemes have been implemented, including
donor cell, van Leer, and piecewise parabolic advection. All results presented
here use the second order van Leer method which has the best
accuracy-to-efficiency
performance.

A fifth step evolves the densities of the separate species according to the
chemistry of the collisional and radiative kinetic equations
\begin{equation}
  \frac{\partial {\rho_i}}{\partial t} = \pm {\sum_{j}}{\sum_{l}}
                                             k_{jl}\rho_j \rho_l
                                         \pm \sum_{j} I_j \rho_j .
\label{split5}
\end{equation}
The methods for solving these equations is the focus of the next subsection.

The total baryonic density $\rho_b$ and the density
of each individual species $\rho_i$ are updated independently in the
expansion, transport and chemistry steps. We are thus able to
monitor the accuracy of our methods by evaluating the constraint
equations for hydrogen, helium and charge conservation, equations
(\ref{conserveH}) -- (\ref{conserveNe}). Over the course of a typical
3D calculation of approximately one-thousand time steps, we
find {\it maximum} errors, which are mostly concentrated at the
shock fronts, to be of order 10 to 30\%. However, for increased
stability and accuracy, we introduce a sixth step in our scheme to
enforce the constraint equations at every time step. An important point
that one must consider in taking this approach is that the errors
can be larger than the concentration of those species that are
depleted. It is therefore necessary to modify only the concentrations of
the dominant species. For a hot gas in which $H$ and $He$
are mostly ionized, the hydrogen and helium constraints should be solved
for the ionized components $H^+$ and $He^{++}$.
In cold ($< 1~eV$) gas the neutral hydrogen and
neutral helium concentrations must
be adjusted to satisfy the constraints. In this way, we are guaranteed to
make only small (typically much less than one percent)
fractional changes to any of the species at each timestep.

\subsection{Solving the Kinetic Equations}
\label{subsec:solving}

More specific details of the numerical methods used in the hydrodynamic
updates (\ref{split1}) -- (\ref{split4}) can be found elsewhere
(Stone \& Norman 1992, Anninos et al. 1994). Here we emphasize the solution
to the chemistry step (\ref{split5}).

Notice that equations (\ref{split5}) can be written schematically as
\begin{equation}
\frac{\partial n_i}{\partial t} = C_i(T,n_j) - D_i(T,n_j)~n_i ,
\label{model}
\end{equation}
where $n_i = \rho_i/(A_i m_H)$ and $A_i$ is the atomic mass number of the
$i$th element and $m_H$ is the proton mass.
The $C_i$ are the collective source terms responsible for the creation of the
$i$th species. The second terms involving $D_i$
represent the destruction mechanisms for the $i$th species and are
thus proportional to $n_i$. Equation (\ref{model}) suggests a 
backward difference formula (BDF) can be used in which all source terms are
evaluated at the advanced time step. Discretization of (\ref{model})
yields
\begin{equation}
n^{t+\Delta t} = \frac{C^{t+\Delta t} \Delta t + n^{t}}
                      {1 + D^{t+\Delta t} \Delta t} .
\label{bdf}
\end{equation}
Lower order backward differentiation methods when applied to problems
of the form $\dot y = f(y,t)$ are stiffly stable. This rather restrictive
stability property is highly desirable when solving sets of stiff equations
(Oran \& Boris 1987).
We have tried other less stable methods including higher order
multi-step predictor-corrector schemes, various
Runge-Kutta and Adams-Bashforth algorithms, and
a Newton's procedure to solve the backwards differenced linearized
equations. All of these alternative
schemes have either proven to be unstable, less accurate
or more expensive computationally compared
to the simple BDF method.

The solver can be optimized further by noting that
the intermediaries $H^-$ and $H_2^+$ in the molecular hydrogen production
processes have large rate coefficients and low
concentrations. They are thus very sensitive to small changes in the
more abundant species. On the other hand, the low concentrations
of $H^-$ and $H_2^+$ implies that they do not significantly
influence the more abundant electron, hydrogen and helium concentrations.
This suggests that the nine species can be grouped into two categories:
fast and slow reacting. The fast reacting group, comprised of $H^-$ and
$H_2^+$,
can be decoupled from the slower network and treated independently since
the kinetic time scales for these species are much shorter
than the characteristic times of the other seven species and
the cosmological or gravitational times. $H^-$ and $H_2^+$
can thus be considered in equilibrium at all times, independent of the
hydrodynamic state variables.

The expressions for the equilibrium abundances of $H^-$
and $H_2^+$ can be reduced by recognizing that reaction (19),
according to Appendix A, can be neglected as a
small order correction to $H^-$, due to the low concentrations
of both species $H^-$ and $H_2^+$. Neglecting reaction (19), the
equilibrium abundance of $H^-$ can be written independent of $H_2^+$
\begin{equation}
n_{H^-} = \frac{k_7 n_H n_e}
     {k_8 n_H + k_{14}n_e + k_{15} n_H + (k_{16}+k_{17}) n_{H^+} + k_{23}} ,
\end{equation}
where the variables $k_i$ are the rate coefficients with subscripts referring
to
the reaction number in Appendix A.
Then given $n_{H^-}$, the equilibrium abundance of $H_2^+$ can
be written with no additional assumptions as
\begin{equation}
n_{H_2^+} = \frac{ k_9 n_H n_{H^+}      + k_{11}n_{H_2}n_{H^+}
                 + k_{17}n_{H^-}n_{H^+} + k_{24}n_{H_2}}
          {k_{10}n_H + k_{18} n_e + k_{19}n_{H^-} + k_{25} + k_{26}} .
\end{equation}
The separation into fast and slow reacting systems helps to further increase
the accuracy and stability of the BDF method when applied only to the slower
network over the longer characteristic time scales required by the Hubble,
hydrodynamic Courant, and gravitational free-fall times in cosmological
simulations.

Due to the intrinsic nonlinearity of equation (\ref{model}), not all
source terms can be evaluated at the advanced time levels. Significant
errors (of order 20\% as measured by the final fractional abundance
of hydrogen molecules) can be introduced if the source terms
$C_i$ and $D_i$ are evaluated at the current time level at
which the species data is known.
Improvements to this crude approximation can be made by
sequentially updating each species in order, rather than updating all species
simultaneously from the data at the past time step.
For example, the order in which we solve the rate
equations was determined finally through experimentation to be
$H$, $H^+$, $He$, $He^+$, $He^{++}$ and $e^-$, followed by the
algebraic equilibrium expressions for $H^-$ and $H_2^+$, then in the
end updating $H_2$ also using the BDF scheme (\ref{bdf}).
The updated concentrations of the $i$th (and previous) species
are used as source terms in the equation for $\dot n_{i+1}$.
Further improvements in accuracy and stability can be made
by mimicking more closely a fully BDF scheme by
subcycling the rate solve
over a single hydrodynamic Courant time step. The subcycle time steps
are determined so that the maximum fractional change in the electron
concentration is limited to 10\% per timestep, ie.
$\delta \tau_{e} = \epsilon n_{e}/\dot{n}_{e}$ with
$\epsilon = 0.1$.

We note that this same subcylcing procedure
can be used to update the energy in the cooling/heating
step in equation (\ref{split2}).
The equation of state (\ref{eqos}) couples the energy and pressure
through the gas temperature, and although we have tried a
Newton-Raphson iterative procedure, we found it to converge very slowly
or sometimes not at all because the cooling/heating rates are strongly
nonlinear and non--monotonic functions of temperature
(Anninos \& Norman 1994).
For this reason we solve equation (\ref{split2}) with an explicit
method that subcycles the cooling source terms. The timesteps for
each subcyle are determined as $\epsilon e/|\dot E - \Gamma|$,
where $\epsilon = 0.1$ as in the rate equation solve.
This algorithm has been tested to be both fast and accurate.

The robustness of our methods has been verified by switching the order in which
the rate equations are solved relative to the other updates. We have also
experimented with the sensitivity to the temperature (ie. time
centered, retarded and advanced temperatures) used in updating
the rate equations. The results are stable and unchanged
under all these sorts of permutations.

\section{CODE TESTS}
\label{sec:codetests}

The numerical scheme described in Section \ref{sec:numerical} has been
implemented in two separate codes: ZEUS-2D and HERCULES. ZEUS-2D is a two
dimensional Eulerian hydrodynamics code originally developed by
Stone \& Norman (1992) and modified for cosmology by
Anninos \& Norman (1994).
HERCULES is a three-dimensional hydrodynamics code derived from a 3D version
of ZEUS, but modified for cosmology and
generalized to include a hierarchical system of nested grids by
Anninos et al. (1994). Extensive tests of the two codes
can be found in the references provided above. In this paper, we present
tests only for the new additions to the codes, namely the multi--species
chemistry and the non--equilibrium cooling.
Due to the intrinsic complexity of such systems, the diversity of
tests is rather limited. We consider two crucial and relevant
(for cosmology) tests:
radiative shock waves and cosmological sheets.
In addition we have also developed
an independent method of solution that is based on the well-tested
packaged solver LSODAR. Fully three-dimensional calculations
of large scale structure formation
are presented, comparing results from our method to that of LSODAR.

\subsection{Radiative Shock Waves}
\label{subsec:radiative}

Chevalier \& Imamura (1982) and Imamura, Wolff \& Durisen (1984)
have demonstrated through both
analytic and numerical calculations that the fundamental mode of oscillations
in one dimensional radiative shocks with cooling rates $\propto \rho^2
T^\alpha$
are unstable if $\alpha < 0.4$.
Because the cooling rate is determined
from the density of electrons and ions, a comparison
to their published work provides an excellent test of both
our cooling algorithm and the reaction network.

The initial data for this test is characteristic of pre-shock flows expected
from the collapse of Zel'dovich pancakes
\begin{eqnarray}
\rho&=&4.72\times 10^{-25}~g~cm^{-3}, \nonumber \\
e   &=&1.0\times 10^{-30}~g~cm^{-1}~s^{-2},    \label{initial_data}  \\
v_x &=&-u_{in}=-1.7\times 10^{7}~cm~s^{-1}, \nonumber
\end{eqnarray}
corresponding to a uniform flow of gas along the $-x$ direction. Reflection
boundary conditions are imposed at $x=0$ and we use 100 zones to resolve
a spatial extent of $L=2.43 \times 10^{-4} Mpc$.
We assume only bremsstrahlung cooling of the form
in Appendix B, which has an exponent of $\alpha \sim 0.5$.

\begin{figure}[t]
\epsfxsize=300pt \centerline{\epsfbox{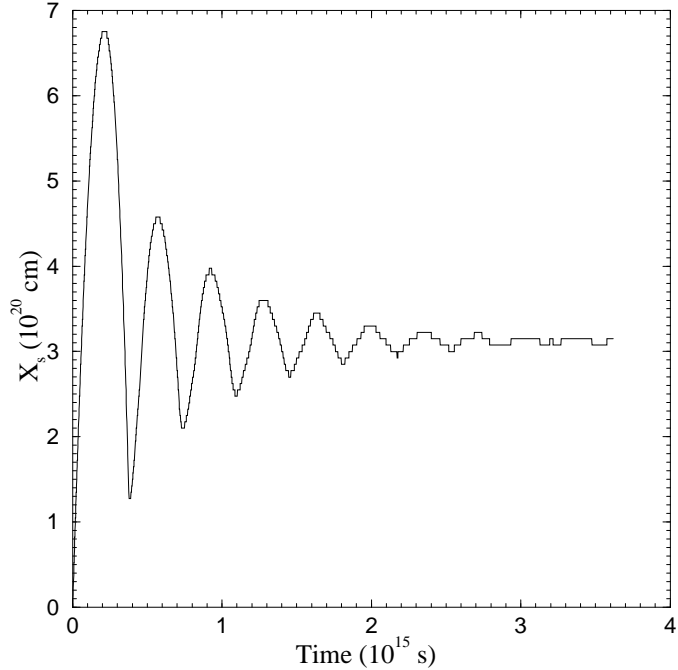}}
\caption{
The location of the shock front $x_s$ as a function of time for the
one dimensional radiative shock test. The complete nine species network
of reactions is solved using the BDF method and a single source of
cooling, Bremsstrahlung. The cooling time, maximum shock position,
period of oscillation, and values of postshock state variables are all
consistent with the analytic and numerical results of
Chevalier \& Imamura (1982) and Imamura et al. (1984).
}
\label{fig:radshk}
\end{figure}

A shock wave forms at the
wall ($x=0$) and propagates outward at a velocity $v_s \sim u_{in}/3$.
As the heated gas cools,
the shock begins to lose pressure support and slows down.
Because the cooling rate is proportional to
$n_e n_{H^+} T^{1/2}$ (ie. $\propto \rho^2 T^{1/2}$),
the cooled gas
experiences an accelerated energy loss as it gets denser.
Eventually the higher density gas nearest the wall loses pressure
support and the shock collapses and
re-establishes a new pressure equilibrium closer to the wall. The shock
front then begins again to propagate away from the wall, repeating
the cycle of oscillations.
This behavior is shown in figure \ref{fig:radshk} where the
shock position $x_s$ is plotted as a function of time.
Our numerical results indicate that the fundamental mode is indeed stable and
damped, consistent with the analytic results of
Chevalier and Imamura (1982) and the numerical simulations of Imamura et al.
(1984).

The shock jump conditions
\begin{eqnarray}
\rho_{post}&=&4\rho_{pre}, \\
p_{post}&=&\frac{4}{3}\rho_{pre}\left(u_{in}\right)^2,
\end{eqnarray}
provide a more quantitative check of our numerical results.
For the choice of initial data (\ref{initial_data}), the jump conditions give
$\rho_{post}=1.9\times 10^{-24}~g~cm^{-3}$ and $p_{post}=1.8 \times
10^{-10}~g~cm^{-1}~s^{-2}$,
in excellent agreement with our numerical results
$\rho_{post}=1.9\times 10^{-24}~g~cm^{-3}$ and $p_{post}=1.7 \times
10^{-10}~g~cm^{-1}~s^{-2}$.

We can estimate the maximum distance the shock front will travel as
\begin{equation}
x_{max} = \tau_{cool} v_s = \frac{e}{\dot E}~v_s ,
\end{equation}
where $\tau_{cool}$ is the cooling time and $v_s$ is the shock speed.
Substituting the bremsstrahlung cooling formula
at the temperature predicted by the jump conditions and assuming
a fully ionized gas with $n_e = n_{H^+} + 2 n_{He^{++}}$ and
hydrogen mass fraction $f_H = \rho_H/\rho_b = 0.76$,
gives $x_{max} \sim 6.5\times 10^{20}~cm$. This is
again consistent with our numerical result $6.7\times 10^{20}~cm$.

Chevalier and Imamura also characterize their linearized analytic solutions
by the frequency of oscillations $\omega_I=0.31$ in units of
$u_{in}/\overline x_s$ where $\overline x_s$ is the average shock front
position. Defining the period $P$ of perturbations as the time for the
shock to first collapse back to the wall, we find $P=3.9\times 10^{14}~s$ and
$\omega_I=(2\pi/P)(\overline x_s/u_{in}) =0.33$, again
in good agreement.

\subsection{Cosmological Sheets}
\label{subsec:sheets}

We use the Zel'dovich (1970) solution to set up a linearized
single mode perturbation for
the collapse of gas in one dimension
\begin{eqnarray}
x    &=& q-\frac{1+z_c}{1+z} \frac{\sin (kq)}{k}, \label{pertx1} \\
v_x  &=&-H_0 (1+z_c)\sqrt{1+z}~\frac{\sin (kq)}{k},  \label{pertv1} \\
\rho &=& \overline\rho \left[1-\frac{1+z_c}{1+z}\cos
(kq)\right]^{-1},\label{pertd}
\end{eqnarray}
where $x$ and $v_x$ are the comoving positions and velocities,
$\rho$ the proper density, $q$ the unperturbed coordinate,
$k=2\pi/\lambda$, $\lambda$ the comoving wavelength of
perturbation, $H_0$ the present Hubble constant,
and $z_c$ the redshift corresponding
to the collapse time.
Parameters for the calculations presented here are the following
\begin{eqnarray}
&& \Omega_0 = 1.0 ,\qquad \Omega_B = 0.04, \qquad  H_0 = 70~km~s^{-1} Mpc^{-1},
   \qquad z_c = 5 ,\qquad \lambda = 10~Mpc  .
\end{eqnarray}

At the higher temperatures ($\gg 1~eV$) characteristic of shock heated gas
in high velocity pancake structures, the kinetic time scales
are extremely short compared to the Hubble time. Neglecting the
photoionization processes, collisional ionization
equilibrium is then a good approximation to the following pairs of reactions

\begin{eqnarray}
\mbox{neutral hydrogen:} \qquad &
  H~+~e^-~&\rightarrow ~H^+~+~2e^- ,  \nonumber \\
& H^+~+~e^-~&\rightarrow ~H~+~\gamma , \label{eq_first} \\
\nonumber \\
\mbox{neutral and singly ionized helium:} \qquad &
  He~+~e^-~&\rightarrow ~He^+~+~2e^- ,\nonumber  \\
& He^+~+~e^-~&\rightarrow ~He~+~\gamma , \\
\nonumber \\
\mbox{negatively charged hydrogen:} \qquad &
  H~+~e^-~&\rightarrow ~H^-~+~\gamma ,\nonumber  \\
& H^-~+~e^-~&\rightarrow ~H~+~2e^- , \\
\nonumber \\
\mbox{ionized molecular hydrogen:} \qquad &
  H~+~H^+~&\rightarrow ~H_2^+~+~\gamma ,\nonumber  \\
& H_2^+~+~e^-~&\rightarrow ~2H , \\
\nonumber \\
\mbox{hydrogen molecules:} \qquad &
  H^-~+~H~&\rightarrow ~H_2~+~e^- ,\nonumber  \\
& H_2~+~e^-~&\rightarrow ~2H~+~e^- . \label{eq_last}
\end{eqnarray}

For a fully ionized gas, we have
\begin{eqnarray}
\frac{\rho_{H^+}}{\rho_b}  & = & f_H , \\
\frac{\rho_{He^{++}}}{\rho_b} & = & 1-f_H ,  \\
\frac{n_e}{n} & = & \frac{n_{H^+}+2n_{He^{++}}}{n_{H^+}+4n_{He^{++}}}
                =   \frac{1+f_H}{2} ,
\end{eqnarray}
where $f_H = 0.76$ is the mass fraction of hydrogen.
The corresponding equilibrium fractional abundances
from reactions (\ref{eq_first}) -- (\ref{eq_last}) can then be written as
\begin{eqnarray}
\frac{n_H}{n}      & = & \frac{k_2}{k_1} f_H ,  \label{eq1} \\
\frac{n_{He}}{n}   & = & \frac{k_4}{k_3} \frac{n_{He^{+}}}{n} , \\
\frac{n_{H^-}}{n}  & = & \frac{k_7}{k_{14}} \frac{k_2}{k_{1}} f_H , \\
\frac{n_{H_2^+}}{n}& = &2\frac{k_9}{k_{18}} \frac{k_2}{k_{1}}
                         \frac{f_H^2}{1+f_H} , \\
\frac{n_{H_2}}{n}  & = &2\frac{k_8}{k_{12}} \frac{k_7}{k_{14}}
                              \left(\frac{k_2}{k_{1}}\right)^2
                              \frac{f_H^2}{1+f_H}
                         \label{eq5}  ,
\end{eqnarray}
which are functions only of the gas temperature.

\begin{figure}[t]
\epsfxsize=300pt \centerline{\epsfbox{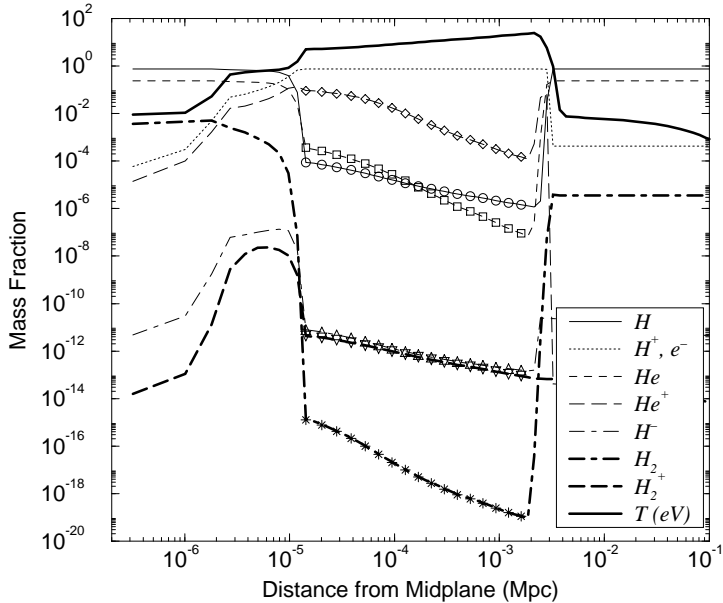}}
\caption{
Mass fractions of the different chemical species at redshift $z=4.8$
across the cosmological sheet which first collapses at $z=5$.
Two distinctive cooling plateaus are evident in the temperature profile
(thick solid line). The gas first cools to $1~eV$ mostly
from atomic hydrogen line cooling
(the region between $3\times 10^{-6} < x < 1\times 10^{-5}~Mpc$)
then cools further (the region $x < 10^{-6}~Mpc$)
due to hydrogen molecules formed from the residual
electron fraction left behind by nonequilibrium effects.
However, collisional equilibrium is a good approximation to the hot gas
between the shock and the outer cooling plateau.
The symbols plotted in this hot region
represent the analytic equilibrium abundances and differ from the numerical
results
by less than 0.5\%.
}
\label{fig:pancake1}
\end{figure}

In figure \ref{fig:pancake1} we show the mass fractions $\rho_i/\rho_b$
throughout the pancake structure at redshift $z=4.8$. At this time,
the shock front is located at a distance of approximately
$3\times 10^{-3}~Mpc$ from the central plane at $x=0$, and is propagating
outward at an average comoving velocity of about 110 $km/s$
(355 $km~s^{-1}$ relative to the infalling gas). Two distinct cooling
layers form, as evidenced by the thick solid line representing the gas
temperature. The gas between $3\times 10^{-6}~Mpc < x < 1\times 10^{-5}~Mpc$
first cools mostly by atomic processes
to a temperature of about $1~eV$. The second colder layer at
$x < 10^{-6}~Mpc$ results from cooling by hydrogen molecules which
form from the residual electrons leftover from the nonequilibrium cooling
through
the first plateau at $1~eV$.
(We refer the reader to Anninos \& Norman (1996) for further details
concerning the chemistry, dynamics and radiative cooling
of cosmological sheets.)
Although nonequilibrium effects are important as the gas cools through
$1~eV$, collisional ionization equilibrium
is a good approximation to the species concentrations in the hot
gas between $2\times 10^{-5}~Mpc < x < 2\times 10^{-3}~Mpc$.
The different symbols plotted across this region in figure \ref{fig:pancake1}
represent the equilibrium abundances given by equations
(\ref{eq1}) -- (\ref{eq5}). Differences between the numerical results
and the equilibrium estimates are less than 0.5\% for all species
except $H_2$ which differ by roughly 5\%. However, the larger discrepancy in
the
$H_2$ mass fraction is not due to numerical errors, but to the
inadequacy of equation (\ref{eq5}) to fully describe the kinetics.
A more accurate equilibrium ratio is derived by considering all the reactions
involving $H_2$
\begin{equation}
\frac{n_{H_2}}{n}   =  \frac{k_8 n_{H^-}n_H +k_{10} n_{H_2^+}n_H
                            +k_{19} n_{H_2^+}n_{H^-}}
                            {k_{11} n_{H^+} + k_{12} n_e + k_{13} n_H}
                       \label{eq6}  .
\end{equation}
Equation (\ref{eq6}) agrees with our numerical results to within about
0.1\%.

\begin{figure}[t]
\epsfxsize=300pt \centerline{\epsfbox{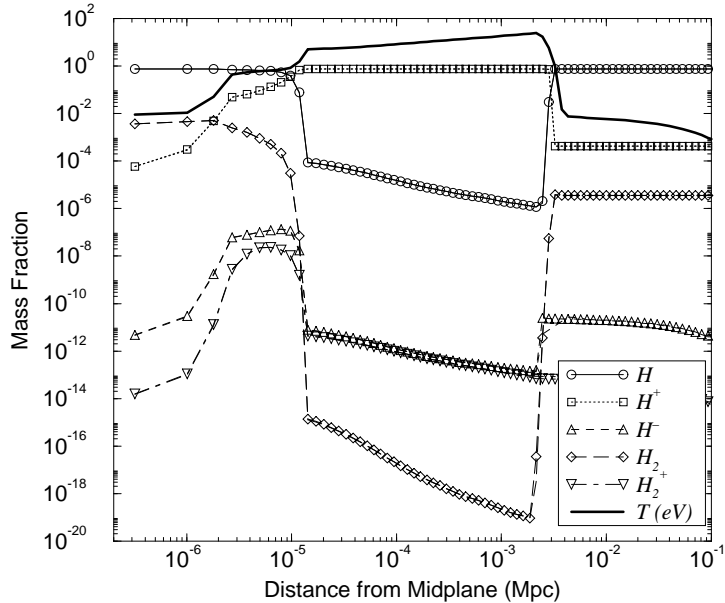}}
\caption{
A comparison of results between the BDF method (symbols) and the
fully non-equilibrium LSODAR algorithms (lines) for the cosmological
sheet of the previous figure, also shown at redshift $z=4.8$.
The two methods agree to within 0.5\% throughout both
the hot and cold pancake layers.
}
\label{fig:pancake2}
\end{figure}

We have also run this same problem using the LSODAR routines in place of
the BDF method to solve for all nine species in full non-equilibrium.
A comparison of the two results at redshift $z=4.8$ is shown in
figure (\ref{fig:pancake2}). The symbols represent the BDF calculation
and the various line types are the LSODAR results for the different
species. The two methods agree to within about 0.5\% throughout all the
different pancake layers, hot and cold. Notice also the excellent
agreement in the mass fractions of $H^-$ and $H_2^+$, which is further
justification of the hybrid model (in which the fast reacting
species are singled out over cosmological dynamical times
to be in equilibrium). Finally we point out that the fractional abundance
of hydrogen molecules that form in the central cooled gas is consistent
with the steady state shock calculations of Shapiro \& Kang (1987).

\subsection{3D CDM evolution}
\label{subsec:cdm}

A final test is presented for an actual three dimensional
cosmological calculation in which
we compare our method of solving the rate equations to that of LSODAR.
Both accuracy and computational efficiency are stressed in this comparison.

The simulation is performed for a flat ($\Omega_0=1$) model universe
with baryon mass fraction $\Omega_{b}=0.035$
and Hubble constant $H_0=100 h~km~s^{-1}Mpc^{-1}$ with $h=0.65$.
The baryonic matter is composed of hydrogen and helium
in cosmic abundance with a hydrogen mass fraction $f_H=0.76$.
The initial data is the Harrison-Zel'dovich power
spectrum modulated with a transfer function appropriate for
cold dark matter (CDM) adiabatic fluctuations
and normalized to a bias factor of $b=1.0$.
We begin the simulations at redshift $z=50$ and evolve to the present
time at $z=0$.
The computational box size is set to $L=128~kpc$ (comoving) resolved
by $32^3$ cells.
Our calculation thus has a spatial grid
resolution of $4~kpc$ and baryonic mass resolution
of $2.6\times 10^2~M_\odot$ which is just marginally adequate to resolve
cooling flows and the formation of hydrogen molecules.
The computational demands of LSODAR implemented in three dimensions
prohibits comparative calculations of much higher resolution.

\begin{figure}[t]
\epsfxsize=300pt \centerline{\epsfbox{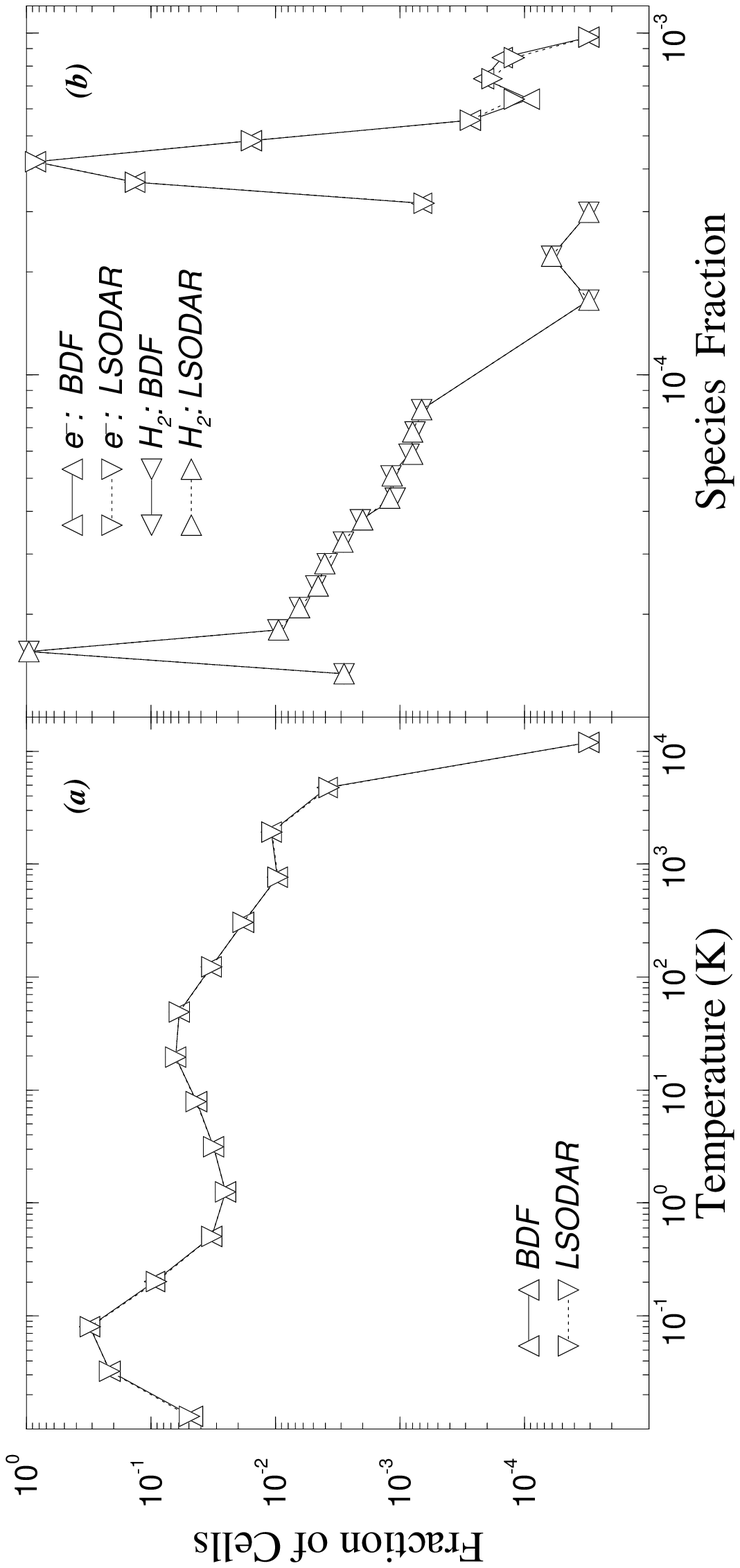}}
\caption{
The cell number distribution is shown for three state variables at redshift
$z=3$. (a) Temperature: The upward and downward pointing
triangles are results from the BDF and
LSODAR solver respectively.
(b) Mass fraction of $H_2$ and the number density fraction of electrons:
The up and down triangles are BDF and LSODAR results for $n_{e}/n$,
while the left and right triangles are BDF and LSODAR
results for $\rho_{H2}/\rho_b$.
The $rms$ differences are 0.018, 0.039, and 0.13 for the
temperature, $H_2$ mass fraction, and electron fraction respectively. The
errors are due in part to the output redshifts not being the same in the 
two cases, and to the large Courant timestep arising from the coarse grid
resolution ($32^3$ cells).
}
\label{fig:hist}
\end{figure}

\begin{figure}[t]
\epsfxsize=400pt \centerline{\epsfbox{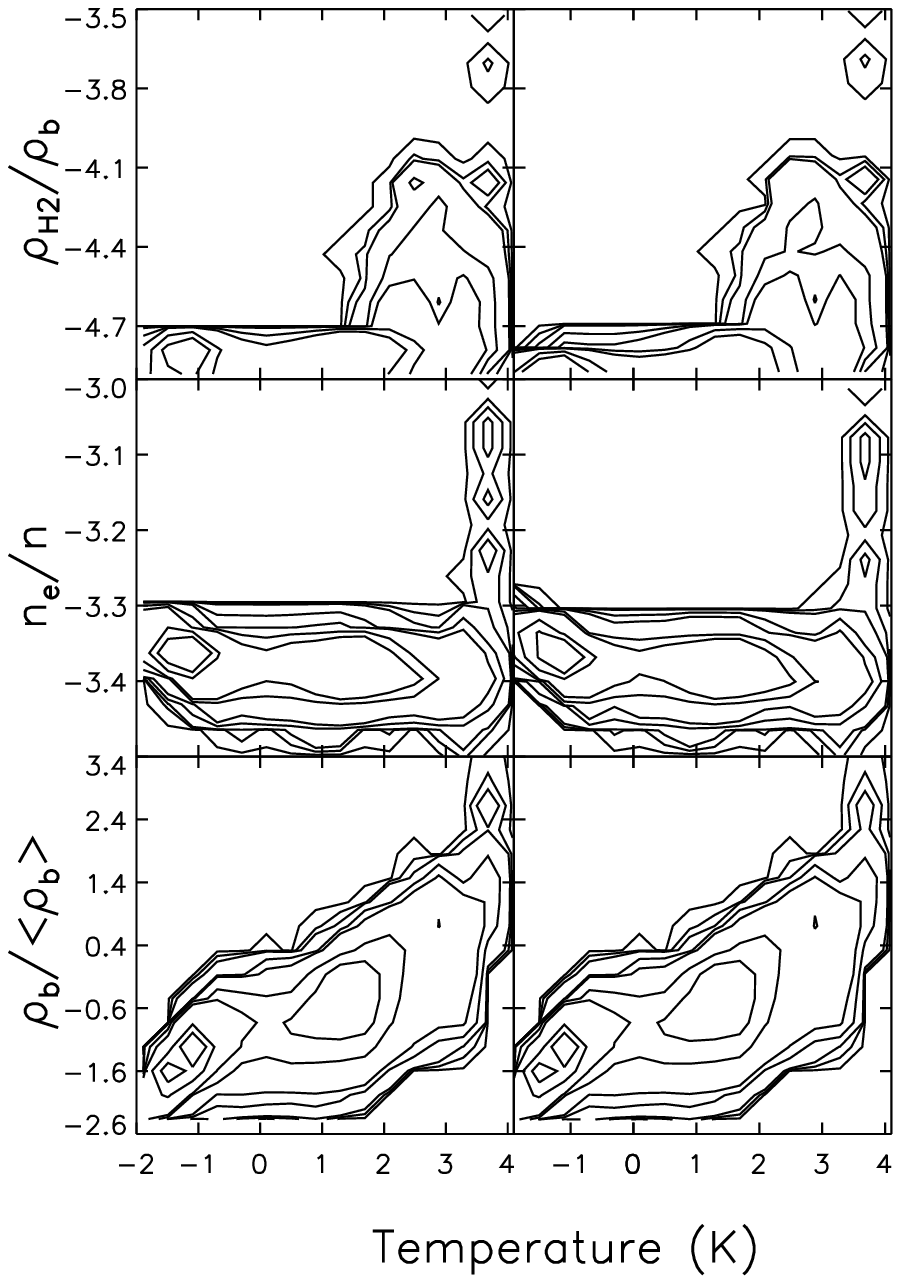}}
\caption{
Contour plots for the cell distribution of the logarithm ($\log_{10}$)
of baryon overdensity,
electron fraction $n_e/n$, and molecular hydrogen fraction $\rho_{H2}/\rho_b$
versus the log of the gas temperature at redshift $z=3$.
Ten contour levels are shown:
($1\times10^{-5}$, $5\times10^{-5}$, $1\times10^{-4}$,
$5\times10^{-4}$, $1\times10^{-3}$, $5\times10^{-3}$, $1\times10^{-2}$,
$5\times10^{-2}$, $1\times10^{-1}$, $5\times10^{-1}$),
which represent the fraction of cells with the specified values of the two
physical quantities.
The three graphs in the left column are results using
the BDF method. The right column are the
LSODAR results. The two methods agree very well.
}
\label{fig:cont}
\end{figure}

\begin{figure}[t]
\epsfxsize=400pt \centerline{\epsfbox{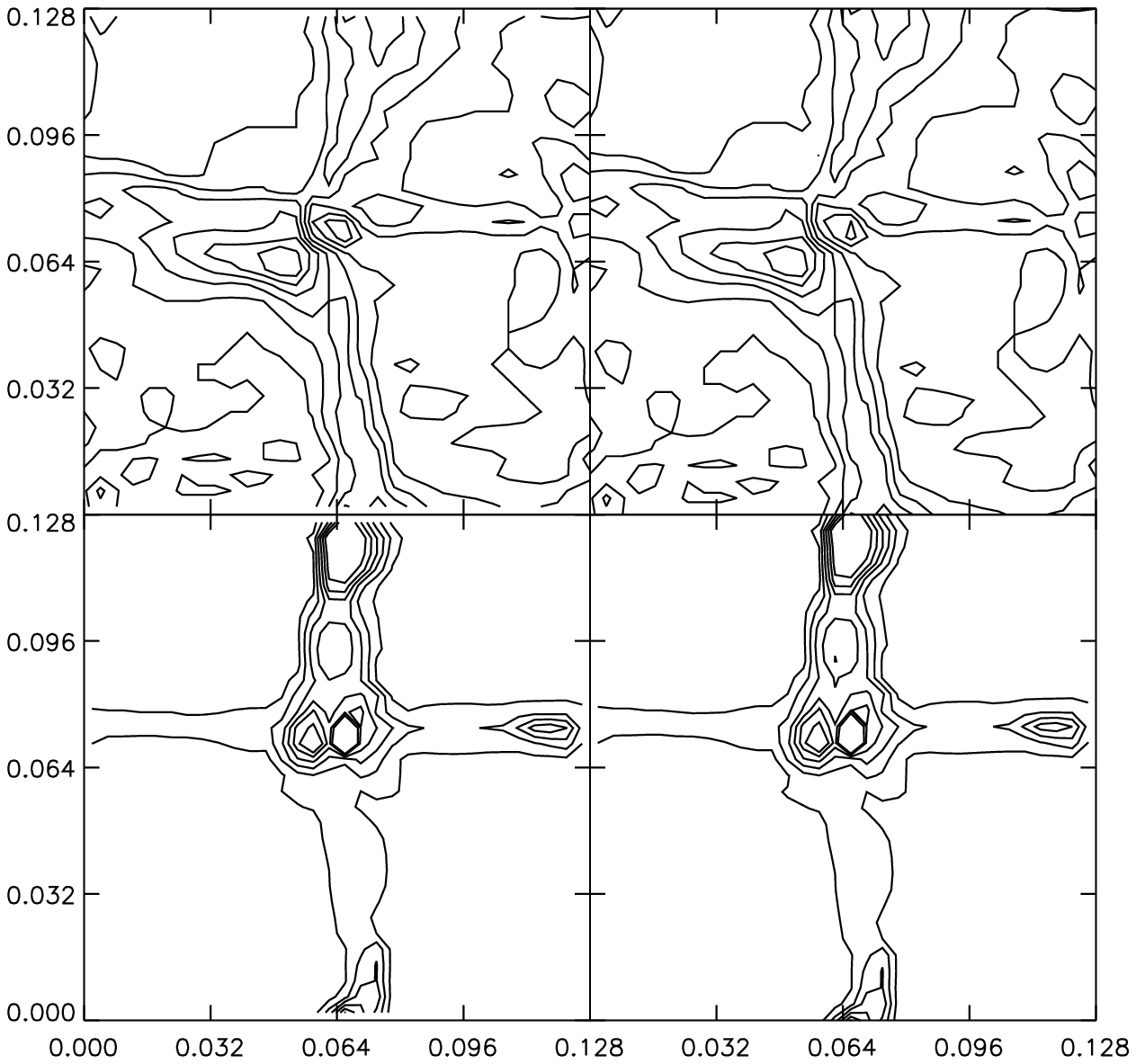}}
\caption{
Contour plots of the spatial distribution of the molecular hydrogen mass and
electron number density fractions at redshift $z=3$.
The lower left panel is $\rho_{H2}/\rho_b$ from the BDF method, and the lower
right
from LSODAR.  The upper left panel is $n_e/n$ from BDF and the upper
right from LSODAR.  The contours cover the entire plane of the computational
box and result from projecting (and averaging) the data along the $z$-axis.
Six contour levels are shown for molecular hydrogen,
(-4.78, -4.75, -4.72, -4.69, -4.66, -4.63), and eight for the electron fraction
(-3.40, -3.39, -3.38, -3.37, -3.36, -3.35, -3.34, -3.33).
Again, in comparing the BDF with the LSODAR results,
the two methods can be seen to agree remarkably well.
}
\label{fig:cont2}
\end{figure}

We have run two separate simulations with identical initial conditions and
model parameters. The only difference in the two runs is the rate equation 
solver: in the one case we use our BDF method, in the other LSODAR.
The results of comparison are shown in figures \ref{fig:hist} to
\ref{fig:cont2}.

Figure \ref{fig:hist} shows the cell distributions (defined by counting up the
number of cells at a particular binned range of values) for three key
variables that would be particularly sensitive to errors in the rate solver:
(a) gas temperature which is constructed from the concentrations
of all the species combined, and
(b) molecular hydrogen mass fraction $\rho_{H2}/\rho_b$
and electron number density fraction $n_e/n$. All results
are shown at redshift $z\sim 3$.
The prominent peaks in the $H_2$ and electron fraction distributions
correspond closely to values initialized at the start of the runs. Because
most of the box volume is comprised of cosmic voids that do not
undergo shock heating, the molecular hydrogen and electron fraction
do not change significantly in most of the cells.
The relative $rms$ deviations between the BDF and LSODAR
results are 0.018, 0.039 and 0.13 for
the gas temperature, molecular hydrogen and electron fractions, respectively.
We note, however, that deviations are due in part to the final
output redshifts not being exactly the same in the two runs; the
relative difference is about 0.01.
The larger differences found here (as compared to the cosmological
sheet calculations) can also be attributed to the coarse grid
resolution and the resulting large Courant time step in the
hydrodynamical calculation. The agreements improve with resolution.

Figure \ref{fig:cont} shows contour graphs of the fractional volume of
those cells with a particular combination of temperature and
baryonic density, electron number density fraction and molecular
hydrogen mass fraction. Two sets of graphs are presented: The
three plots in the first column
are results from the BDF method, the second column are the LSODAR results.
We note that the sharp boundaries in the electron contours at the level
$\log_{10} (n_e/n) \sim -3.3$ correspond to the initialized fraction
at $z=50$: $n_e/n = 1.2\times 10^{-5} (h\Omega_b)^{-1}$ (Peebles 1993).
The volume weighted distribution is concentrated at the initial value for
$H_2$ since there is no mechanism to efficiently create nor destroy molecules
at the
low temperatures of the voids, and slightly lower than the initial value for
the electron fraction since the expansion (cosmological and gravitational)
of the voids continues to cool and recombine the gas.

Contours of the spatial distribution of $n_e/n$ and $\rho_{H2}/\rho_b$
are shown in figure \ref{fig:cont2}. The data is projected (and averaged)
along the $z$-axis at redshift $z=3$.
The first row is the electron fraction, the second molecular hydrogen.
The first column are the BDF results, the second LSODAR.
Notice that hydrogen molecules form preferentially within the high dense
filamentary
structures but mostly in the knot--like intersections of the filaments.
These are the highest density regions where the gas cools rapidly and
the electrons are depleted by recombination.
Peaks in the $H_2$ concentration therefore correspond
to valleys in the electron distribution.
In comparing results from the two solvers, BDF versus LSODAR,
we see the distributions
in both figures \ref{fig:cont} and \ref{fig:cont2} are basically the same.

In addition to the accuracy of the BDF method, another very important point
that should be stressed is the amount of computational time required
to solve the rate equations.
The BDF run takes about 1.2 CPU hours with the full reaction
network on the NCSA Convex C3880, in contrast to the equivalent
LSODAR calculation which takes about 16.7 hours
to complete on the same machine.  The speedup of the BDF method over LSODAR
is roughly a factor of 14.

\section{Summary}
\label{sec:summary}

We have developed and tested a new scheme to solve a system of stiff
kinetic equations appropriate for chemical reaction flows in cosmological
structure
formation. Twenty--eight chemical reactions for collisional and radiative
processes
are included in our model, which tracks nine separate
atomic and molecular species:
$H$, $H^+$, $He$, $He^+$, $He^{++}$, $H^-$, $H_2^+$, $H_2$, and $e^-$.
The reaction network is solved in a self-consistent manner with the
hydrodynamic, N-body and cosmological expansion equations,
and the accuracy of the solver has been verified by
performing a series of test-bed calculations that includes
radiative shock waves, cosmological sheets and monitering the conservation
constraints. We have also implemented
a publicly available and well-tested
solver called LSODAR in place of our
scheme and made direct comparisons of the different results
in one, two and three dimensions. We find our methods are
both fast and accurate, making fully
three-dimensional calculations of non-equilibrium
cosmological reaction flows feasible.

We have incorporated  the species solver into two separate cosmological
hydrodynamic codes: a two--dimensional ratioed grid code to model Zel'dovich
pancakes,
and a more general three--dimensional nested grid cosmological code HERCULES.
Applications to date include investigations of star and galaxy formation
in cosmological sheets (Anninos \& Norman 1996), primordial star formation
in CDM models (Zhang et al. 1996), and simulations of the
Ly$\alpha$ forest (Zhang, Anninos \& Norman 1995; Charlton et al. 1996).
In the future, we plan to extend this work and
develop a more sophisticated treatment of radiation
to account for self--shielding and to more accurately model
the microphysics of optically thick gas.

\acknowledgments
\centerline{\bf ACKNOWLEDGEMENTS}
\vskip19pt

This work is done under the auspices of the Grand Challenge Cosmology
Consortium (GC$^3$) and supported in part by NSF grant ASC-9318185.
The calculations were performed on both the C90 at the Pittsburgh
Supercomputing
Center and the Convex C3880 at the National Center for Supercomputing
Applications
at the University of Illinois.

\newpage
\centerline{\bf APPENDIX A: Chemical Reactions}
\vskip19pt

The following is a list of all chemical reactions that we include in our
calculations. Further discussions and justification of the
completeness of this set of reactions can be found in Abel et al. (1996).
There we also include explicit formulae for the different rate coefficients.

\begin{tabbing}
\hskip55pt \= (16)\hskip10pt \= $He^{++}$ \= + \hskip10pt $H^{+}$
                             \= $\rightarrow$
                             \hskip10pt $He^{++}$  \= +   \hskip10pt $H^{+}$
                             \hskip15pt \=  \kill
\hskip35pt {\bf Collisional Processes:} \\
\hskip55pt \> (1)\hskip10pt \> $H$ \> + \hskip10pt $e^{-}$   \> $\rightarrow$
                               \hskip10pt $H^+$    \> + \hskip10pt 2$e^{-}$
\hskip55pt \>  \\
\hskip55pt \> (2)\hskip10pt \> $H^+$ \> + \hskip10pt $e^{-}$ \> $\rightarrow$
                               \hskip10pt $H$      \> + \hskip10pt $\gamma$
\hskip55pt \>  \\
\hskip55pt \> (3)\hskip10pt \> $He$  \> + \hskip10pt $e^{-}$ \> $\rightarrow$
                               \hskip10pt $He^{+}$ \> + \hskip10pt 2$e^{-}$
\hskip55pt \>  \\
\hskip55pt \> (4)\hskip10pt \> $He^+$\> + \hskip10pt $e^{-}$ \> $\rightarrow$
                               \hskip10pt $He$     \> +  \hskip10pt $\gamma$
\hskip55pt \>  \\
\hskip55pt \> (5)\hskip10pt \> $He^+$\> + \hskip10pt $e^{-}$ \> $\rightarrow$
                               \hskip10pt $He^{++}$\> + \hskip10pt 2$e^{-}$
\hskip55pt \>  \\
\hskip55pt \> (6)\hskip10pt \> $He^{++}$\> + \hskip10pt $e^{-}$ \>
$\rightarrow$
                               \hskip10pt $He^+$ \> + \hskip10pt $\gamma$
\hskip55pt \>  \\
\\
\hskip55pt \> (7)\hskip10pt \> $H$   \> + \hskip10pt $e^{-}$ \> $\rightarrow$
                               \hskip10pt $H^-$    \> +  \hskip10pt  $\gamma$
\hskip55pt \>  \\
\hskip55pt \> (8)\hskip10pt \> $H^-$   \> + \hskip10pt $H$\> $\rightarrow$
                               \hskip10pt $H_2$   \> +  \hskip10pt  $e^{-}$
\hskip55pt \>  \\
\hskip55pt \> (9)\hskip10pt \> $H$   \> + \hskip10pt $H^+$\> $\rightarrow$
                               \hskip10pt $H_2^+$ \> +  \hskip10pt  $\gamma$
\hskip55pt \>  \\
\hskip55pt \> (10)\hskip10pt \> $H_2^+$ \> + \hskip10pt $H$\> $\rightarrow$
                                \hskip10pt $H_2$  \> +  \hskip10pt  $H^+$
\hskip55pt \>  \\
\hskip55pt \> (11)\hskip10pt \> $H_2$ \> + \hskip10pt $H^+$\> $\rightarrow$
                                \hskip10pt $H_2^+$\> +  \hskip10pt  $H$
\hskip55pt \>  \\
\hskip55pt \> (12)\hskip10pt \> $H_2$ \> + \hskip10pt $e^{-}$  \> $\rightarrow$
                                \hskip10pt $2H$   \> +  \hskip10pt  $e^{-}$
\hskip55pt \>  \\
\hskip55pt \> (13)\hskip10pt \> $H_2$ \> + \hskip10pt $H $ \> $\rightarrow$
                                \hskip10pt $3H$  \>
\hskip55pt \>  \\
\hskip55pt \> (14)\hskip10pt \> $H^-$ \> + \hskip10pt  $e^{-}$  \>
$\rightarrow$
                                \hskip10pt $H$    \> + \hskip10pt  2$e^{-}$
\hskip55pt \>  \\
\hskip55pt \> (15)\hskip10pt \> $H^-$ \> + \hskip10pt $H$  \> $\rightarrow$
                                \hskip10pt $2H$   \> +  \hskip10pt  $e^{-}$
\hskip55pt \>  \\
\hskip55pt \> (16)\hskip10pt \> $H^-$ \> + \hskip10pt $H^+$\> $\rightarrow$
                                \hskip10pt $2H$   \>
\hskip55pt \>  \\
\hskip55pt \> (17)\hskip10pt \> $H^-$ \> + \hskip10pt $H^+$\> $\rightarrow$
                                \hskip10pt $H_2^+$\> + \hskip10pt  $e^{-}$
\hskip55pt \>  \\
\hskip55pt \> (18)\hskip10pt \> $H_2^{+}$\> + \hskip10pt $e^{-}$\>
$\rightarrow$
                                \hskip10pt $2H$  \>
\hskip55pt \>  \\
\hskip55pt \> (19)\hskip10pt \> $H_2^{+}$\> + \hskip10pt $H^-$\> $\rightarrow$
                                \hskip10pt $H_2$  \> +  \hskip10pt $H$
\hskip55pt \>  \\
\\
\hskip35pt {\bf Radiation Processes:} \\
\hskip55pt \> (20)\hskip10pt \> $H$ \> + \hskip10pt $\gamma$ \> $\rightarrow$
                                \hskip10pt $H^+$  \> + \hskip10pt $e^{-}$
\hskip55pt \>  \\
\hskip55pt \> (21)\hskip10pt \> $He$\> + \hskip10pt $\gamma$ \> $\rightarrow$
                                \hskip10pt $He^+$ \> + \hskip10pt  $e^{-}$
\hskip55pt \>  \\
\hskip55pt \> (22)\hskip10pt \> $He^+$ \> + \hskip10pt $\gamma$\> $\rightarrow$
                                \hskip10pt $He^{++}$ \> + \hskip10pt  $e^{-}$
\hskip55pt \>  \\
\hskip55pt \> (23)\hskip10pt \> $H^-$  \> + \hskip10pt $\gamma$\> $\rightarrow$
                                \hskip10pt $H$   \> +  \hskip10pt  $e^{-}$
\hskip55pt \>  \\
\hskip55pt \> (24)\hskip10pt \> $H_2$  \> + \hskip10pt $\gamma$\> $\rightarrow$
                                \hskip10pt $H_2^{+}$ \> + \hskip10pt  $e^{-}$
\hskip55pt \>  \\
\hskip55pt \> (25)\hskip10pt \> $H_2^+$\> + \hskip10pt $\gamma$\> $\rightarrow$
                                \hskip10pt $H$   \> + \hskip10pt  $H^+$
\hskip55pt \>  \\
\hskip55pt \> (26)\hskip10pt \> $H_2^+$ \> + \hskip10pt $\gamma$\>
                                $\rightarrow$
                                \hskip10pt $2H^+$\> + \hskip10pt  $e^{-}$
\hskip55pt \>  \\
\hskip55pt \> (27)\hskip10pt \> $H_2$  \> + \hskip10pt  $\gamma$\>
                                $\rightarrow$ \hskip10pt $H_2^*$ $\rightarrow$
                                \hskip10pt $2H$    \>
\hskip55pt \>  \\
\hskip55pt \> (28)\hskip10pt \> $H_2$  \> + \hskip10pt  $\gamma$\>
                                $\rightarrow$
                                \hskip10pt $2H$    \>
\hskip55pt \>  \\
\end{tabbing}

\newpage
\centerline{\bf APPENDIX B: Cooling and Heating Processes}
\vskip19pt

The cooling rates $\dot E$
and photoionization cross sections $\sigma_j (\nu)$ included in our
calculations. We use units of $ergs~cm^{-3} s^{-1}$
for the rates, $cm^{2}$ for the cross sections, and degrees Kelvin for
temperature $T$. Also, $T_n = T/10^n$,
$\alpha_i=\sqrt{(\nu/\nu_{0,i})-1}$ with $\nu_{0,i}$
being the threshold frequencies of the $i$th species,
and $k_1$, $k_3$ and $k_5$
are the rate coefficients for the ionizing chemical reactions
(1), (3) and (5) listed in Appendix A.

\begin{tabbing}
\hskip15pt \=
     $7.5\times10^{-19}(1+(T/{10^5})^{1/2})e^{-118348/T}~n_e n_{H}$ \kill
{\bf Collisional excitation cooling} (Black 1981; Cen 1992) : \\
\hskip15pt \>  $7.50\times10^{-19}~(1+\sqrt{T_5})^{-1}~
                \exp{(-118348/T)}~n_e n_{H}$\\
\hskip15pt \>  $9.10\times10^{-27}~(1+\sqrt{T_5})^{-1}~
                T^{-0.1687}\exp{(-13179/T)}~n_e^2 n_{He}$ \\
\hskip15pt \>  $5.54\times10^{-17}~(1+\sqrt{T_5})^{-1}~
                T^{-0.397}~\exp{(-473638/T)}~n_e n_{He^+}$ \\
\\
{\bf Collisional ionization cooling} (Shapiro \& Kang 1987; Cen 1992) : \\
\hskip15pt \>  $2.18\times 10^{-11}~k_1 n_e n_{H}$ \\
\hskip15pt \>  $3.94\times 10^{-11}~k_3 n_e n_{He}$ \\
\hskip15pt \>  $8.72\times 10^{-11}~k_5 n_e n_{He^+}$ \\
\hskip15pt \>  $5.01\times10^{-27}~(1+\sqrt{T_5})^{-1}~
                T^{-0.1687}~\exp{(-55338/T)}~n_e^2 n_{He^+}$ \\
\\
{\bf Recombination cooling} (Black 1981; Spitzer 1978) : \\
\hskip15pt \>  $8.70\times10^{-27}~T^{1/2}~T_3^{-0.2}~
                (1+T_6^{0.7})^{-1}~n_e n_{H^+}$ \\
\hskip15pt \>  $1.55\times10^{-26}~T^{0.3647}~n_e n_{He^+}$ \\
\hskip15pt \>  $1.24\times10^{-13}~T^{-1.5}~[1+0.3\exp{(-94000/T)}]~
                \exp{(-470000/T)}~n_e n_{He^+}$ \\
\hskip15pt \>  $3.48\times10^{-26}~T^{1/2}~T_3^{-0.2}~
                (1+T_6^{0.7})^{-1}~n_e n_{He^{++}}$ \\
\\
{\bf Molecular hydrogen cooling} (Lepp \& Shull 1983). \\

\\
{\bf Bremsstrahlung cooling} (Black 1981; Spitzer \& Hart 1979): \\
\hskip15pt \>  $1.43\times10^{-27}~T^{1/2}
                ~[1.1+0.34\exp{(-(5.5-\log_{10} T)^2/3)}]
                ~n_e (n_{H^+} + n_{He^+} + n_{He^{++}})$ \\
\\
{\bf Compton cooling or heating} (Peebles 1971) :  \\
\hskip15pt \>  $5.65\times10^{-36}~(1+z)^4~[T-2.73(1+z)]~n_e$ \\
\\
{\bf Photoionization heating crossections} (Osterbrock 1974) : \\
\hskip15pt \>  $\sigma_{H}(\nu)=6.30\times10^{-18}~(\alpha_H^2+1)^{-4}~
                [\exp{(4-4 (\tan^{-1}~\alpha_H)/\alpha_H)}]~
                [1-\exp{(-2\pi/\alpha_H)}]^{-1}$ \\
\hskip15pt \>  $\sigma_{He}(\nu)=7.42\times10^{-18}~
                [1.66(\alpha_{He}^2+1)^{-2.05}
                -0.66(\alpha_{He}^2+1)^{-3.05}]$ \\
\hskip15pt \>  $\sigma_{He^+}(\nu)=1.58\times10^{-18}~(\alpha_{He^+}^2+1)^{-4}~
                [\exp{(4-4(\tan^{-1}~\alpha_{He^+})/\alpha_{He^+})}]~
                [1-\exp{(-2\pi/\alpha_{He^+})}]^{-1}$ \\
\end{tabbing}


\end{document}